\documentclass[toc]{PoS}

\usepackage{graphicx}
\usepackage{lineno}
\usepackage{enumerate}
\usepackage{cite}
\usepackage{wrapfig}

\title{Power-aware applications for scientific cluster and distributed
  computing}
\ShortTitle{}

\author{David Abdurachmanov$^a$, \speaker{Peter Elmer}$^b$, Giulio Eulisse$^c$, Paola Grosso$^d$, Curtis Hillegas$^e$, Burt Holzman$^c$, Ruben L. Janssen$^d$, Sander Klous$^d$, Robert Knight$^e$, Shahzad Muzaf{}far$^c$ \\
\llap{$^a$} Digital Science and Computing Center, Faculty of Mathematics and Informatics, Vilnius University, Vilnius, Lithuania \\
\llap{$^b$} Department of Physics, Princeton University, Princeton, New Jersey 08540, USA \\
\llap{$^c$} Fermilab, Batavia, IL 60510, USA \\
\llap{$^d$} University of Amsterdam, Kruislaan 904, Amsterdam, The Netherlands \\
\llap{$^e$} Research Computing, Office of Information Technology, Princeton University, Princeton, New Jersey 08540, USA \\

E-mail: \email{Peter.Elmer@cern.ch}
}

\abstract{

The aggregate power use of computing hardware is an important cost
factor in scientific cluster and distributed computing systems. 
The Worldwide LHC Computing Grid (WLCG) is a major example of such
a distributed computing system, used primarily for high throughput
computing (HTC) applications. It has a computing capacity and power
consumption rivaling that of the largest supercomputers. The 
computing capacity required from this system is also expected to
grow over the next decade. Optimizing the power utilization and cost of 
such systems is thus of great interest.

A number of trends currently underway will provide new opportunities
for power-aware optimizations. 
We discuss how power-aware software applications and scheduling might be 
used to reduce power consumption, both as autonomous entities and as 
part of a (globally) distributed system. As concrete examples of computing 
centers we provide information on the large HEP-focused Tier-1 at 
FNAL, and the Tigress High Performance Computing Center at Princeton
University, which provides HPC resources in a university context.}

\FullConference{International Symposium on Grids and Clouds (ISGC) 2014,\\
       23-28 March 2014\\
       Academia Sinica, Taipei, Taiwan}

\begin{document}

\section{Introduction}
\label{sec:intro}
  Over the past 15 years the use of globally distributed computing 
resources has been a critical ingredient for many large scientific
projects. For example, the Worldwide LHC Computing Grid (WLCG)~\cite{WLCG} 
has been built to serve the needs of the high energy physics (HEP) experiments 
at the Large Hadron Collider (LHC)~\cite{LHCMACHINE} at the European Laboratory
for Particle Physics (CERN) in Geneva, Switzerland. Numerous scientific 
results have been produced using the WLCG, including in particular the 
discovery of the Higgs Boson~\cite{CMSHIGGS,ATLASHIGGS}, by the 
CMS~\cite{CMSDET} and Atlas~\cite{ATLASDET} experiments, which led
to the 2013 Nobel Prize in Physics. 

  The WLCG today brings together computing resources from nearly 160 
computer centers in 35 countries, with 
approximately 350,000 x86 cores of compute power and 200PB of storage, 
with 10Gb network links connecting most centers. Although it is not 
traditionally considered as a supercomputer, the compute capacity of the 
WLCG resources is similar to those provided by the most powerful 
supercomputers~\cite{TOP500}. Just as importantly, our rough estimates
(scaling up from the set of machines currently in use at FNAL)
of the power requirements of the WLCG are
O(10MW), even before taking into account the Power Usage Ef{}fectiveness 
(PUE) of the participating computer centers. This is again in line with
the most powerful supercomputers. 

  Planning is also underway for a major upgrade, the High Luminosity Large 
Hadron Collider (HL-LHC)~\cite{HLLHC}, which will run through $\sim$2030. 
The data volume increase over the next 15 years will be O($10^3$) and naive 
extrapolations from today's software indicates that compute needs could 
increase by factors O($10^3$-$10^4$), significantly faster than usual
Moore's Law extrapolations. Significant ef{}{}forts will be required to reduce
that to a level which is possible within funding limits, but it is likely 
that significant globally contributed compute resources will be continue to
be required.

  In this paper, we focus in particular on the power aspects of this
large distributed computing system. The distributed ownership of the 
clusters which make up the WLCG implies that the resulting costs are also 
distributed, however optimizations are in principle possible at all
levels. For a given set of hardware, there are three possible goals
for throughput-oriented scientific computing applications:

\begin{enumerate}[(a)]
    \item optimizing for maximum performance or throughput (the primary focus today), and/or
    \item optimizing to reduce the total power use for fixed throughput, i.e.\ the power efficiency (throughput per Watt), and/or
    \item optimizing to reduce total power cost for a given throughput
\end{enumerate}
We will refer back to these three goals in the later sections of this
note.

\section{Computing Model - Current Practice}
\label{sec:cmpresent}

  The construction of the WLCG was greatly facilitated by the convergence, 
around the year 2000, on Linux and commodity x86 processors as the standard 
for HTC scientific clusters like those used in HEP.
The resulting homogeneity significantly simplified the use of the compute 
resources as a ``build once, run anywhere'' scheduling model was possible 
for the application software. 

  The distributed computing and data management models of HEP experiments,
however, introduced a simplifying restriction whereby ``jobs are sent to 
data''. The computing model of the CMS experiment~\cite{CMSCTDR} is a typical 
example. 
Datasets are transferred in advance and placed statically in storage
(e.g.\ with tools like PhEDEx~\cite{PHEDEX}) in one or more centers. 
Each time it is necessary to run an application
using data from those datasets as input, the relevant application software
is transferred to one of the centers where the required input data 
can be read via LAN access from site-local storage (``data locality''). In 
order to insure the full utilization of compute resources replicas of data 
are made manually in multiple sites as required. In the first years 
of the WLCG many of the computer centers were new and not yet operating at 
full reliability. The static placement strategy minimized dependencies
between centers and allowed for ef{}ficient and scalable commissioning
of the resources.  It had however the disadvantage of requiring
significantly more storage than strictly necessary due to the dataset
replication. The wide area network (WAN) was also underutilized as
a resource, despite being significantly more robust than originally
imagined.

The applications used in HEP are embarrassingly parallel, with no
communication required between running instances. Each application
simply reads some piece of input data and outputs as a result some
processed or reduced version of that data. The applications themselves
have a simple, sequential design (no threading). This was consistent
with the decades long trend by industry to turn exponential increases
in transistor counts over time for the same cost (Moore's Law)
into equivalent exponential gains in the performance of sequential
software applications like those used in HEP. 
Around 2005, power density limitations in chips ended this trend,
and sequential
applications can no longer be made to run exponentially faster on
subsequent generations of processors~\cite{GAMEOVER}. This is true
even if the underlying transistor count continues to increase per
Moore's Law.  The initial response of industry to these power density
limits was the introduction of ``multi-core'' CPUs, with more than one
functional processor on a chip.  
HEP responded to the evolution from
single-core to multi-core by simply scheduling
additional sequential applications onto the additional cores. In this
way net throughput continues to scale roughly with Moore's Law even
if individual
application performance does not.
As a downside, this has greatly increased the number of simultaneous
running jobs in the computing systems, increased the total amount of memory
needed and placed more stringent requirements on the scalability of job
management systems, data access systems, and so forth.

While aggregate power costs and limitations have been important in 
individual centers, the WLCG system as a whole does not account for or use 
information about power use or costs operationally, nor do the 
systems or software used by the WLCG users (i.e.\ HEP experiments).

\section{Computing Model - Evolution}
\label{sec:cmfuture}

  The computing model described in Sec.~\ref{sec:cmpresent} represents the
period in which the WLCG was created through the end of the first
LHC Run ($\sim$2006-2013). The computing model is now evolving, or
under pressure to evolve, for a variety of reasons. In this section
we go through the most important changes and argue that, taken together,
they enable an operational model where the WLCG can be treated as an integrated, global ``power-aware'' system.

\subsection{Transition to multi-core aware software applications} 
\label{sec:multicore}
   As described earlier, the introduction of multicore
processors in $\sim$2005 was an epochal change. Power (density) limitations
ended the era of Moore's Law scaling for sequential applications, as multi-core
processors were introduced. Since the available applications were
still sequential (non-threaded), an ``application per core'' scheduling
model was adopted. As more cores appear in processors, this model led
to the use of batch schedulers to schedule ever smaller fractions of the 
total computing resources (within batch nodes, and within individual CPU dies).

 HEP experiments are now in the process of transitioning to multi-core
aware (threaded) application frameworks~\cite{FWKMC,FWKMCR,GAUDIMC} which
are capable of scheduling work across multiple cores. In the near term,
this will resolve some current problems with the ``application per core'' 
model, such as ever increasing total memory requirements and scheduler 
scaling issues with the number of total jobs. In the medium term, however, 
it represents a significant opportunity for optimization of the processor use,
both for performance and for power, at the application level. Although
programming multi-threaded applications is more complex than sequential 
applications, they can potentially make choices and coordinate resources
for performance and power efficiency better than coarse-grained, incoherent 
scheduling of independent processes by a batch scheduler. The newer
model will be hierarchical: the batch scheduler will schedule a
larger resource (e.g.\ a whole node) and within that the software
application will schedule local resources. 

For various data management, workflow or bookkeeping reasons, it is also
often the case that individual jobs are configured to process a
minimum amount of input data (e.g.\ an entire file or data 
taken during a fixed amount of time such as a ``luminosity section'').
This minimum ``job size'' often implies a minimum wall clock time
duration for jobs which can at times be many hours or, in extreme
cases, days. Before 2005 increasing processor clock frequencies
reduced these minimum job durations over time. In the multicore
era, only multi-threaded applications can reduce wall clock times
for such jobs. Jobs which complete in a shorter wall clock time,
but nonetheless use many processor cores, are useful when power costs
vary over the day.
 


\subsection{Processor Technology}
\label{sec:cputech}
The same power density limitations that led to multicore CPU's are
having additional effects. The performance-per-cost improvement of 
40\% or more per year, seen in the 1990's and early 2000's~\cite{MLGROWTH}, 
has been replaced by more meager expectations of 20-25\% per
year~\cite{CERNSURVEY}, even when taking into account the full use of multicore processors.
In the worst case, this corresponds to a factor of 6 growth 
instead of a factor of $\sim$30 from Moore's Law over the next decade. The
need for larger performance gains drives interest in the use of other 
processor architectures, including ``lightweight'' low-power
general purpose cores, 
specialized architectures like GPUs and/or large vector units, etc. 
These architectures typically have better performance per watt than
today's standard general purpose cores, however fully exploiting these newer 
architectures typically implies significant reengineering of software 
applications~\cite{GAMEOVER,SMPROC}.

Examples of these alternate architectures are already being deployed today,
in the form of coprocessors (GPUs and Intel's Xeon Phi), in computer
centers shared with other scientific fields. HEP applications are
in general not yet capable of using them, however many preliminary studies 
have been done~\cite{GPU4,HALYO1,FUNKE,pw_chep2013,pw_nss2013,GPU5,GPU3,HAUTH} and a few specific applications are leading the way~\cite{GPU7,GOOFIT,madgraph_gpu,ICECUBE}. Several groups have investigated the use of low power mobile 
processors (e.g.\ ARM) for HEP 
applications~\cite{ACAT13ARM, CHEP13ARMPHI, CHEP13LHCBARM}.
In fact there is some expectation that the power density limitations will
lead to System-on-Chip (SoC)~\cite{SC13HPCSOC} architectures mixing
special purpose GPUs with lower power general purpose cores.
The net effect of this evolution will be significantly more heterogeneity in 
the computing systems in use, with large variations in performance
and power 
characteristics both within a given center and between centers. In 
addition significant variation in the efficiency with which applications
use any given hardware configuration is likely.




\subsection{Data Federations} 

The reliability of all WLCG computer centers has greatly improved
from the experience gained during the first LHC run. 
More sophisticated data management and access models are thus
possible. In particular it is now possible to relax the ``data locality''
requirement. For example, CMS is currently deploying a worldwide 
data federation~\cite{AAACHEP13} using the xrootd data access system to 
access data remotely across the WAN~\cite{XROOTD1, XROOTD2}.
Via the data federation an application running in one center can
open a file for reading, and the system will find and allow remote
reads from a copy of the file wherever it is located in the world.
Efficient remote access to data removes compute and storage locality
requirements, and allows for many additional job scheduling options for
throughput or power efficiency reasons.


\subsection{WLCG as a global power-using computing system} 

  As described above, the WLCG is evolving. The introduction of
multi-threaded applications and the decoupling of locality requirements
for storage add significant new flexibility to global job scheduling. At
the same time, evolution towards heterogeneous computing hardware and
heterogeneous resource provisioning will require more sophisticated
approaches to achieve both maximum throughput and power efficient
usage.
As such, it becomes possible to consider the WLCG not only as a globally
distributed computing system, but also as a system with significant
power requirements and many opportunities for optimization of the power use.

%

\section{Existing Research on Energy Efficiency}


Optimization of power costs for large computing centers is of general
interest and significant research has been done.
Already in 2007 energy proportional computing~\cite{ENERGYPROP} was 
identified as an emerging focus area for data centers and large scale 
infrastructures. The goal in data centers was and is threefold, namely one 
wants to:
\begin{itemize}
\item decrease consumption in order to achieve lower operating costs and
reduce environmental impact
\item maintain performance by exploiting better energy proportionality of hardware components
\item schedule applications in an energy optimal manner.
\end{itemize}
Reducing the power costs of large scale infrastructures can be
accomplished in several manners. \cite{PGBBB} and \cite{PGCCC} were among 
the first to
propose to exploit the differences in energy pricing in order to place
location of computation in a power aware manner. The knowledge of the
actual power profiles of the equipment in the data centers and of the
network devices connecting them allows to distribute computation and
data storage more efficiently, as we have shown in~\cite{PGDDD}.

Exploiting hardware characteristics requires proper models and energy
profiles of hardware components. Work in this area abounds:~\cite{PGEEE}
focused on modeling virtual machine's contribution to the consumption of
physical nodes;~\cite{PGFFF,PGGGG} proposed statistical power models for 
GPUs;~\cite{PGHHH} profiles network equipment.
Application scheduling and resource scheduling are emerging now as an
integrated approach for large scale distributed systems~\cite{PGIII}.

  In the High Performance Computing (HPC) area, there has been much
focus in recent years on Exascale Computing~\cite{DOEEXAPROG,DOEEXATECH}, including the challenge of
limiting the total power consumption of such machines to O(20MW).
Exascale computing has some points in common with HTC, in
particular regarding the evolution of hardware. However it differs
in two important ways. First, synchronization requirements of large 
parallel HPC applications place more constraints on runtime scheduling 
choices and
use of power saving functionalities (e.g.\ dynamic voltage and frequency
scaling, DVFS) than HTC applications, which can be deployed independently
in a fine-grained fashion. Second, whereas an
HPC installation will be located in a single location with a single hardware
architecture, distributed HTC (as in the WLCG) allows for exploitation of 
geographically heterogeneous power costs and heterogeneous hardware choices.

In this note, we explore the application of some of these ideas to
distributed HTC systems, taking into account the specific
characteristics of these kinds of applications.

\section{Example Computer Centers}
\label{sec:centers}

  No full compilation of the energy use
of all of the WLCG centers exists at this time. Here we
provide some illustrative examples of relevant characteristics for two
large computing centers.

\subsection{Princeton University Tigress High Performance Computing Center}
\label{sec:picscie}
The Tigress High Performance Computing Center at Princeton University
is an example of a university research computing facility
used for a broad mix of scientific and engineering research purposes
including both traditional HPC and batch-oriented high throughput
applications. Although not currently integrated into WLCG, it is
used as a Tier3 resource by CMS.

The dedicated computer center was built with the intention of replacing
the research computing clusters set up by individual departments
and research groups in university buildings not designed for
that purpose. The center opened in 2011 on the Princeton Forrestal
campus. It was designed for up to 5MW utility power into the building
and 3MW to the data hall at full build out (currently only 1.8MW of
UPS is provisioned). 
To supplement the utility power, it has a 2.5MW emergency diesel generator
and a 2.0MW natural gas generator. The latter is used at times that
electricity is expensive, currently 1800+ hours per year. It is designed
for a PUE of 1.5 at full build out, and currently varies 1.4-1.7.
Typical power use today is 1.2MW.

Power costs range from $\sim$\$0.08 to \$0.80 / kW hour in New Jersey,
with an average of \$0.09 - \$0.10 / kW-hour. Princeton pays on a minute
by minute basis. The co-generation plant on campus includes predictive
modeling to know when to turn on and off. The natural gas generator
leverages this capability to know when to turn on and off.
While not all electricity markets have this kind of dynamic pricing,
in places where it exists it provides for work scheduling choices to
reduce power cost.


%

\subsection{FNAL Tier 1 Computing Center}
\label{sec:fnal}
\begin{figure}[tbp]
  \centering
  \includegraphics[width=0.6\textwidth]{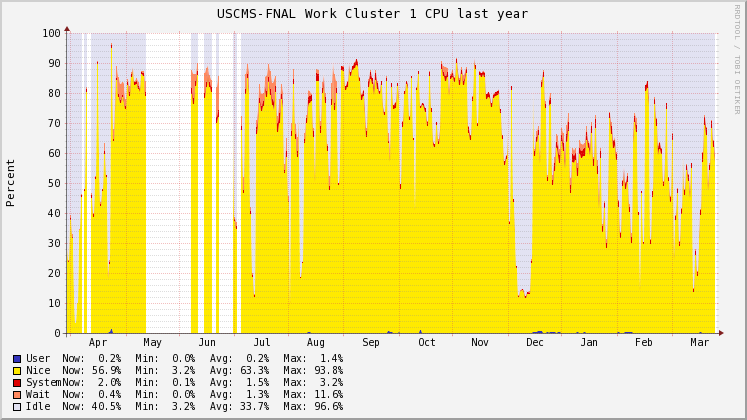}
  \caption{Average CPU utilization of nodes at FNAL Tier1 site.}
  \label{fig:fnalcpueff}
\end{figure}

\begin{figure}[tbp]
  \centering
  \includegraphics[width=0.6\textwidth]{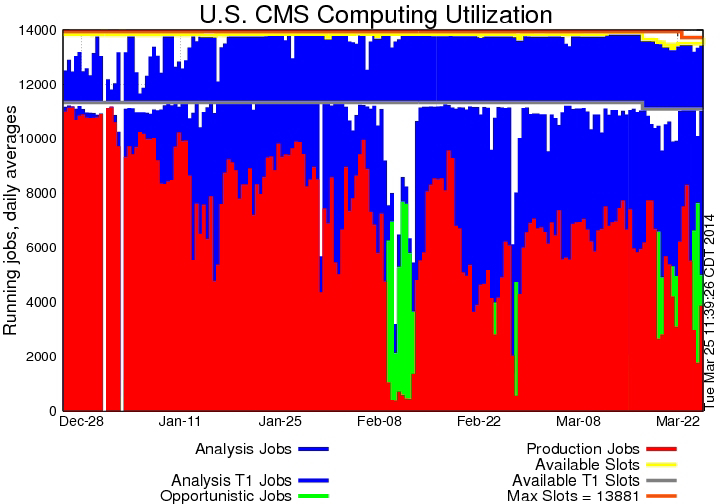}
  \caption{Job slot (core) utilization, by job type, for FNAL Tier1 site.
The lower plot is for the Tier1 grid cluster, the upper plot is for the
LPC analysis cluster.}
  \label{fig:fnalcpuutil}
\end{figure}

  Fermi National Accelerator Laboratory (FNAL), in Batavia, IL (USA)
provides Tier-1 computing center resources as part of the CMS experiment.
Fig.~\ref{fig:fnalcpueff} shows the average CPU utilization efficiency
over the past year for the set of nodes comprising the WLCG Tier1
computing center. The average CPU efficiency varies over time and in
practice is often less than 100\%. Fig.~\ref{fig:fnalcpuutil} shows
the utilization by job type for both the Tier1 and for a dedicated
analysis cluster. Two things are visible. First, utilization of all
available job slots is high, but not always 100\%. At times
some slots will be idle. Second, the origin of the variation in
CPU utilization efficiency is clearer: jobs can be classified 
into two rough categories: ``production'' jobs with high CPU
efficiency and ``analysis'' jobs with lower efficiency. The analysis
jobs typically need to read more data (relative to computation)
and are waiting on input data.
(The change in job mix can in part be attributed to the fact that 
early 2014
corresponds to the midpoint of a 2 year shutdown of the LHC for upgrades,
and thus a reduced need for ``production'' jobs and a higher fraction
of analysis jobs.)
The type characteristics of jobs are often known or can be measured,
as 100's or 1000's of jobs of a given type are usually submitted
together. Again, work scheduling choices can reduce power cost.

\section{Computing Hardware}

The power consumption of individual hardware components, in
particular processors, is one important ingredient for the
optimization of overall power use and cost. Most WLCG sites have
multiple purchases of computing nodes made over the past 4-5 years,
with some reporting nodes in use for up to 8 years after purchase.
Different
generations of nodes having wildly different performance per unit
power characteristics.  (Up to a factor of 20 in the case of 8
years of purchases.) In addition, the future introduction of
either
low-power cores or co-processor cards (GPUs, Xeon Phi), coupled
with varying application efficiency in using them, will introduce further
heterogeneity.

\begin{figure}[tbp]
  \centering
  \includegraphics[width=0.6\textwidth]{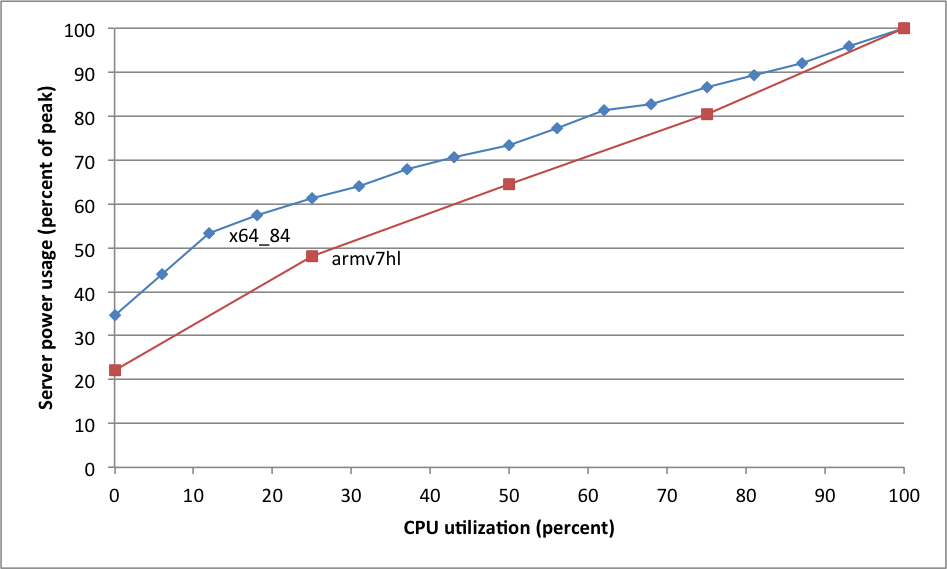}
  \caption{ARMv7 (Exynos 5410, A15, performance governor, 1.6Ghz) Efficiency,
           x86-64 (Intel(R) Xeon(R) CPU E5-2670 0 @ 2.60GHz) Efficiency}
  \label{fig:powerarm}
\end{figure}

When jobs are not 100\% efficient an additional important aspect 
is the power proportionality of the systems. Fig.~\ref{fig:powerarm}
shows one such comparison, between a standard x86 server and (for
illustration) a simple development board for a low-power processor.
In cases where CPU utilization is less than 100\% due to job
inefficiency or due to empty job slots (idle cores), these curves
illustrate the varying cost of different machines. Also in this
case, the introduction of co-processors will change significantly
these curves, and their importance will grow as application efficiency
in using the co-processors may vary significantly. For hardware
combinations that are not fully power-proportional (i.e.\ consuming
zero power for zero throughput), the most power-efficient working
point is of course 100\% utilization, but this is not always
achievable.

Note that the non-zero idle power contains contributions not only
from the (idle) processor, but also from the motherboard, power
supply, internal disks and the bulk memory. HTC clusters are often
built to ``worst case'' application requirements and typically extra
memory and/or local disk capacity is added because of small
purchase-time cost. As power use and cost become more important,
characterization of power proportionality curves at purchase time
will also help quantify eventual operational costs from such choices.

\section{Performance-Aware Applications and Scheduling}


Before considering the power-related aspects of applications and
scheduling, we first consider additional emerging opportunities from
performance-aware applications and scheduling, i.e.\ goal (a)
in Sec.~\ref{sec:intro}.

An application's performance can be characterized by its throughput,
CPU efficiency, memory use, etc.
Today's simple, sequential applications are however typically not
designed to measure anything about their performance because they were not designed with any significant
run-time adaption to change their behavior.
At best they make such measurements, but purely for reporting
purposes. All ``adaption'' is usually left to external entities such as
the workflow management system or even to the user/programmer to
adapt in the next version of the program.

As described in Sec.~\ref{sec:multicore} and~\ref{sec:cputech}
application designs are evolving to use multiple cores and 
specialized processing capabilities such as GPUs, 
explicitly taking on work- and resource-scheduling responsibilities
that sequential applications lacked. These designs in principle can
evolve to measure performance and adapt the amount and type of work
they schedule within the application to maximize throughput. Typically
they do this today in a basic fashion at job startup, e.g.\ by noting
the number of cores and/or GPUs available and configuring the job to 
use them. Similar optimizations are possible for memory and local disk
use for caches. A logical next step would be for the job to adapt itself
as it runs based on performance measurements, e.g.\ by starting additional
threads, scheduling larger chunks of input data, etc. This kind of 
fine-grained performance optimization cannot be done by a batch 
scheduler, but only by the software application itself.

Performance measurements from the application itself, especially
in terms of ``average throughput'' achieved, can also be used for
more effective job scheduling. (We use {\it throughput},
meaning
average output per unit time, and not absolute {\it output}, as the
latter
will vary depending on particular job configurations.) Already
today, sequential application performance will vary from (general
purpose) core to core depending on processor type (CPU, cache memory,
etc.). How it varies will also depend on the specifics of the
particular application type. Note that it should be possible for
each application type to measure ``throughput'' on its own terms.
Rather than use the absolute value of that quantity for scheduling
purposes, suitably normalized (across processor type) averages can
be calculated and the marginal advantage of any particular scheduling
choice could in principle be calculated and comparisons made between
different application types. This becomes even more interesting as 
hardware becomes heterogeneous and application designs permit running on 
multiple hardware configurations (e.g.\ code which runs via OpenCL on
both CPUs and GPUs). In this way, scheduling choices can be made to 
maximize the overall throughput of the heterogeneous system with
a mix of application types.

\section{Power-Aware Computing}

  Once optimizations to achieve maximum throughput have been made,
a logical next step would optimization of power use or cost for
a given throughput, i.e. goals (b) and (c) from Sec.~\ref{sec:intro}.
We consider a number of cases.

In the simple case where there is insufficient work in the queue
(i.e.\ job slots are empty, as is seen from time to time in
Fig.~\ref{fig:fnalcpuutil}), two possibilities exist. The simplest
solution is that new jobs can be scheduled on the most
power efficient hardware possible. For general purpose CPUs,
the power efficiency can simply be determined via one-time hardware characterization and used for all subsequent scheduling choices. For
more complex heterogeneous mixes or varying CPU utilization efficiencies,
power-aware applications might be needed to provide both throughput
and power use characterization of each application type. Measuring
the power requires both hardware and operating system support (e.g.\ 
via interfaces such as IPMI). A more complex solution would be 
aggregation (via scheduling or virtual machine migration)
of all jobs on the most power efficient nodes and (temporary) powering
down unused nodes if their idle power is non-negligible.

In the more common case where a sufficient queue of work exists, but
the applications have a mix of processor utilization efficiencies,
two possibilities exist. Processor utilization efficiencies can
be less than 100\% for multiple reasons, including latencies due 
to input and output (I/O) of data, insufficient parallel computation (for
multi-threaded or codes which use coprocessors) and/or insufficient
memory. As in the previous section one can still define a throughput
metric for each application type on its own terms, and similarly
a throughput per power metric, and use these to do job scheduling.
Note that this naturally encompasses situations where a particular
application cannot use a coprocessor card such as a GPU, but is able
to use the host CPUs on the same machine. The gain in terms of 
relative throughput for each application type for scheduling choices as 
well as idle power is treated quantitatively.

In the presence of process inefficiency and when sufficient memory is
available, one strategy today to increase the aggregate throughput
is to overschedule the compute resource with multiple jobs. Some sites
in fact do this, however it is limited by the available memory and
in cases where I/O is limited by filesystem access rather than latencies
it may be counterproductive. In addition, overscheduling of emerging
architectures (larger numbers of lower power cores, coprocessor cards) 
is more complex than general purpose cores.

Thus the two methods for exploiting a mix of application efficiencies
both aim to reduce power cost, but exploiting temporal or geographical
price variations in power cost. As described in the Tigress example
in Sec.~\ref{sec:picscie}, power costs can vary dynamically in some
centers and some component of this variation is predictable (e.g.\ 
day/night or peak/off-peak use variations). The temporal power-aware
scheduling choice within a single center would be to schedule the
most processor efficient jobs (with the highest power needs) 
during the periods of lowest power cost, and those with lower processor
utilization efficiency during the periods of higher cost. Net throughput
would in principle remain the same on average. Similarly, a geographical
power-aware scheduling choice is available on distributed computing
systems like the WLCG which takes advantage of the fact that the 
phase of temporal variation and average local power costs vary from center
to center.

\subsection{Simulation results}

\begin{figure}[tbp]
  \centering
  \includegraphics[width=0.6\textwidth]{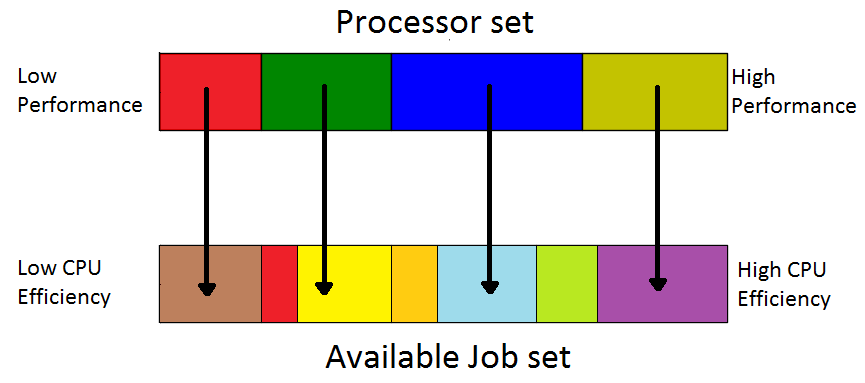}
  \caption{Scheduling scheme in the WLCG simulation: processor from
           the available processor sets are mapped to jobs in the available
           job set.}
  \label{fig:rj1}
\end{figure}

A simple simulation~\cite{RJTHESIS} allowed us to quantify the relation
between possible performance improvements and energy savings in the
WLCG under different scheduling policies. We focused on seven Tier2
sites, for which we knew the processor types present and the job
history from information in the CMS Dashboard~\cite{DASHBOARD}.
Fig.~\ref{fig:rj1} shows the scheduling scheme we simulated; we
relied on two sets, a processor set and an available job set, to
couple jobs and processor with each other. The processor set
distinguishes available nodes in terms of their performance. The
job set relies on information from actual WLCG job logs to classify
incoming jobs as low and high CPU efficient based on the author
previous running history.

\begin{figure}[tbp]
  \centering
  \includegraphics[width=0.6\textwidth]{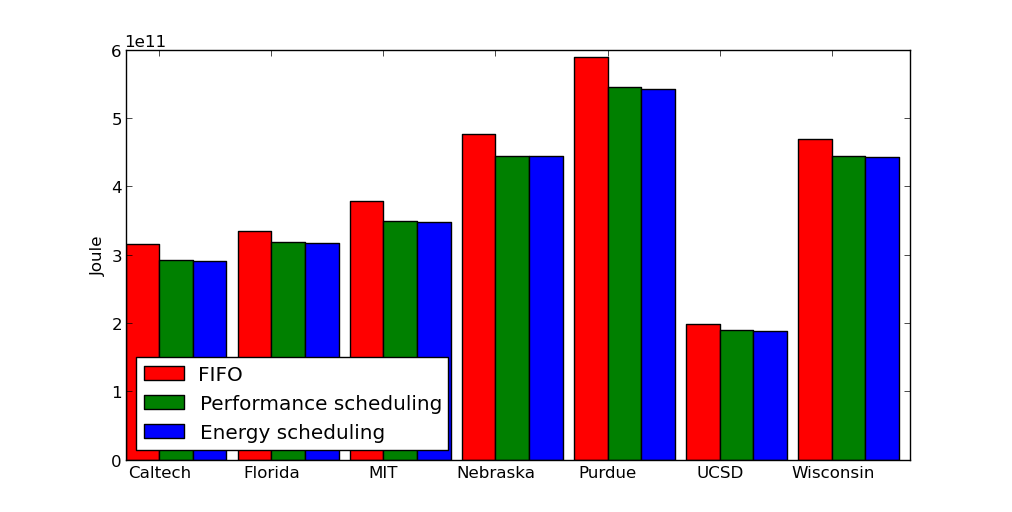}
  \caption{Energy consumption per site under three different scheduling 
          algorithms.}
  \label{fig:rj2}
\end{figure}

\begin{figure}[tbp]
  \centering
  \includegraphics[width=0.6\textwidth]{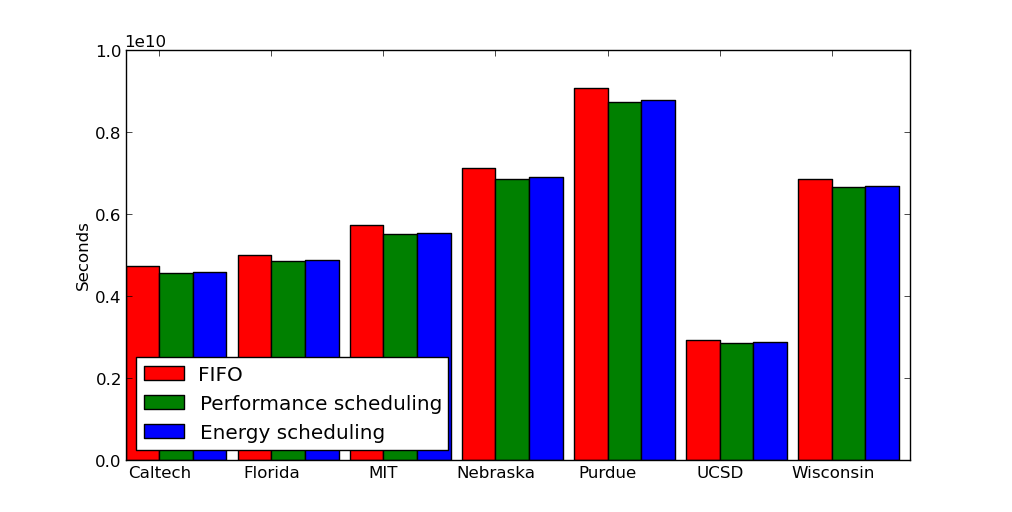}
  \caption{Performance per site under the three different algorithms.}
  \label{fig:rj3}
\end{figure}

Once a processor becomes available the scheduling algorithm will
find the best job in the job pool for the specific processor. For
our experiments we used three different types of algorithms:
(1) FIFO scheduling selects the first initiated job in the queue
is executed first;
(2) Energy efficient scheduling aims to schedule CPU intensive jobs
on relatively energy efficient processors. This results in lower
total energy consumption as CPU intensive jobs benefit more than
the lesser  CPU intensive jobs of the better energy profile of the
processor; and
(3) CPU performance scheduling schedules jobs to processors with a
performance that is proportional to the estimated relative CPU
efficiency of the job.
Fig.~\ref{fig:rj2} shows that both the performance and energy
scheduling significantly reduce energy consumption in comparison
to FIFO scheduling; the average improvements over the seven sites
are respectively 6.74\% and 7.07\%. Similarly Fig.~\ref{fig:rj3} shows
an improvement in the performance of the two efficiency-oriented
algorithms over the basic scheduler. The average improvement in
this case is 3.44\% for the CPU scheduling and 2.95\% for the energy
scheduling.

These results don't identify energy or performance scheduling to
be best, but clearly show that they should be preferred to job and
processor agnostic schedulers. Given the focus in Tier2 sites is
on performance the CPU performance scheduler will be in most cases
the optimal choice.

%
%

\section{Conclusions and Future Work}

In the preceding sections, we have described general ideas as to
how power optimizations could be made on a large distributed
computing system like the WLCG. We take into account current and
future computing models and foreseen technology evolution.


The first step we are taking is to model distributed systems such as the
WLCG that present different power cost functions, heterogeneous hardware
choices and serve a diverse range of applications.
Our upcoming work will proceed in steps toward this goal. First we will 
determine what it is possible to measure today at each of the sites in 
terms of power use, as well as possible variations the power costs in WLCG 
computer centers. Secondly, this macro knowledge will be coupled with further
characterization of the evolving power and performance characteristics
of available hardware, including coprocessors and SoCs. Ultimately by 
determining what node/processor level power savings might be possible through 
application choices, we will be able to develop power aware applications on 
heterogeneous hardware and we will examine power and throughput impact 
of power-aware scheduling.


  In this paper we have discussed the application and exploitation
of various techniques to reduce power consumption and cost in
distributed high throughput computing clusters such as the WLCG.
Such techniques will be of use to build the computing systems of
ever greater computational throughput required for the scientific
problems of the next decade and beyond.

\section*{Acknowledgments}
This work was partially supported by the National Science Foundation, under
Cooperative Agreement PHY-1120138, and by the U.S. Department of Energy.

\end{document}